\documentstyle[12pt]{article}
\advance \textheight by \topskip
\begin{document}

\begin{titlepage}
\hbox to \hsize{\hfil IHEP 96--12}
\hbox to \hsize{\hfil hep-ph/9602392}
\hbox to \hsize{\hfil March, 1996}
\vfill
\large \bf
\begin{center}
Hyperon Polarization in the Constituent Quark Model
\end{center}
\vskip 1cm
\normalsize
\begin{center}
{\bf S. M. Troshin and N. E. Tyurin}\\[1ex]
{\small  \it Institute for High Energy Physics,\\
  Protvino, Moscow Region, 142284 Russia}
\end{center}
\vskip 1.5cm
\begin{abstract}

We consider
mechanism for hyperon polarization  in inclusive
production. The main role belongs to the orbital angular momentum
and polarization of
the strange quark-antiquark pairs in the internal structure of
the constituent quarks.
We consider a nucleon
as a core consisting of the constituent quarks embedded into quark
condensate.
The nonperturbative hadron structure is
based on the results of chiral quark models.
\end{abstract}
\vfill
\end{titlepage}

\section*{Introduction}

One of the most puzzling and persistent since a long time spin effect
was observed in inclusive hyperon production in collisions of
unpolarized hadron beams. A very significant polarization of
$\Lambda$--hyperons has been discovered two decades ago \cite{hell}.
Since then  measurements in different processes
were performed \cite{newrev} and number of models was
proposed for qualitative and quantitative description of these  data
\cite{xxx}. Among them the Lund model based on classical string mechanism
of strange quark pair production \cite{and}, models based on spin--orbital
interaction \cite{miet} and multiple scattering of massive strange sea
quarks
in effective external field \cite{szw} and also models for polarization of
$\Lambda$ in diffractive processes with account for proton states with
additional $\bar{s} s$ pairs such as $|uud\bar{s}s\rangle$ \cite{trt,khar}.
It was proposed also to connect $\Lambda$ polarization in the process
$pp\rightarrow\Lambda X$ with the polarization in the process
$\pi p\rightarrow \Lambda K$ \cite{soff} and use triple Regge approach
\cite{prep}.

The mechanism of gluon fusion in perturbative QCD
as a source of strange quark
polarization has been considered in \cite{gold}
and $x$ and $p_\perp$--dependencies of
$\Lambda$--polarization has been discussed.

Nevertheless, hyperon polarization phenomena are
 not completely
 understood in   QCD and currently could be considered
 even as a more serious
problem than the problem of proton spin which hopefully will find its final
resolution in near future.  Of course, those problems are interrelated  and
one could attempt to connect the spin structure of nucleons studied in
deep--inelastic
scattering with the polarization of $\Lambda$'s observed in
hadron production.
As it is widely known now, only part (less
than one third in fact) of the proton spin is due to quark spins
\cite{ellis,altar}.  These results can be interpreted in the
effective QCD approach ascribing a substantial part of hadron spin
to an orbital angular momentum of quark matter.  It is natural to
guess that this orbital angular momentum might be revealed in
asymmetries in hadron production.

It is also evident from deep--inelastic scattering data
\cite{ellis,altar,vos}
that
strange quarks play essential role in the  proton structure and in
its spin balance in particular. They are negatively polarized
in a polarized nucleon, $\Delta s\simeq -0.1$.
Polarization effects in  hyperon production also continue to
 demonstrate \cite{newrev} that strange
 quarks produced in hadron interactions appear to be polarized.

In the recent papers
\cite{asy} we
considered a possible origin of asymmetry in the pion and
$\varphi$--meson
production under collision of a polarized proton beam with unpolarized proton
target and argued that the orbital angular momentum of partons inside
constituent quarks  leads to significant asymmetries in meson
production.
In this paper we consider
how the most characteristic features of hyperon and
first of all $\Lambda$ polarization can be accounted in such approach.

\section{Structure of constituent quarks}

 We consider a nonperturbative hadron as consisting of the
  constituent quarks located at the central part of the hadron which
embedded into a quark condensate.  Experimental and theoretical
arguments in favor of such a picture were given,  e.g. in
\cite{isl,trtu}.  We refer to effective QCD and the use the NJL model
\cite{njl} as a basis. The  Lagrangian  in addition to the
four--fermion interaction of the original NJL model includes
the six--fermion $U(1)_A$--breaking term.

 Transition to partonic picture
 in this model is described by the introduction of a momentum cutoff
 $\Lambda=\Lambda_\chi\simeq 1$ GeV, which corresponds to the scale
of chiral symmetry spontaneous breaking.  We adopt
the point that the need for such cutoff is an effective implementation of the
short distance behaviour in QCD \cite{jaffe}.

The constituent quark
masses can be expressed in terms of quark condensates \cite{jaffe}, e.g.:
\begin{equation} m_U = m_u-2g_4\langle 0|\bar u u|0\rangle-2g_6\langle 0|\bar
d d|0\rangle \langle 0|\bar s s|0\rangle .\label{ms} \end{equation} In this
approach massive  quarks appear  as quasiparticles, i.e. as current quarks
and the surrounding  clouds of quark--antiquark pairs which consist of a
mixture of quarks of the different flavors.
It is worth to stress that in
addition to $u$ and $d$ quarks constituent quark ($U$, for example) contains
 pairs of strange quarks (cf. Eq. (\ref{ms})).
Quantum numbers of the
constituent quarks are the same as the quantum numbers of current quarks due
to the conservation of the corresponding currents in QCD.  The only exception
is the flavor--singlet, axial--vector current, it has a $Q^2$--dependence
due to axial anomaly which arises under quantization.

Quark radii are determined by the radii of  the
clouds surrounding it.  We assume that the strong interaction radius
 of  quark  $Q$  is determined by its Compton wavelength:
 $r_Q=\xi /m_Q$, where
 constant $\xi$ is universal for different  flavors. Quark
 formfactor $F_Q(q)$ is taken in the dipole form, viz
\begin{equation} F_Q(q)\simeq (1+\xi^2{\vec{q}}^{\,2}/m_Q^2)^{-2}
\label{ff} \end{equation} and the corresponding quark matter
distribution $d_Q(b)$ is of the form \cite{trtu}:  \begin{equation}
d_Q(b)\propto \exp(-{m_Qb}/{\xi}). \label{bf} \end{equation}

   Spin of constituent quark $J_{U}$  in this approach is given
 by the  following sum \begin{eqnarray}
 J_{U}=1/2 & = & J_{u_v}+J_{\{\bar q q\}}+\langle L_{\{\bar qq\}}\rangle=
\nonumber\\
              & & 1/2+J_{\{\bar q q\}}+\langle L_{\{\bar qq\}}\rangle.
\label{bal}
\end{eqnarray}
The value of the orbital
 momentum contribution into the spin of constituent quark can be
 estimated with account for new experimental results from
 deep--inelastic scattering \cite{vos} indicating that quarks
 carry even less than
 one third of proton spin, i.e.
\[
(\Delta\Sigma)_p\simeq 0.2,
\] and taking into account  the relation between
 contributions of current quarks into a proton spin and corresponding
contributions of current quarks into a spin of constituent quarks
and that of constituent quarks into  proton spin \cite{altar}:
\begin{equation} (\Delta\Sigma)_p = (\Delta U+\Delta
D) (\Delta\Sigma)_U.\label{qsp} \end{equation}
If we adopt that  $\Delta U+\Delta D=1$
   \footnote{We will use this simplest
assumption, which is enough for our estimates. However, account of orbital
   and gluonic effects at the level of constituent quarks reduces $\Delta
   U+\Delta D$ by 25$\%$ \cite{lip,kiss}.}
then we should conclude that   $J_{u_v}+J_{\{\bar q q\}}=1/2(\Delta\Sigma)_U
\simeq 0.1$ and from
Eq.  (\ref{bal}) $\langle L_{\{\bar q q\}}\rangle\simeq 0.4$, i. e. about
   80\% of the $U$ or $D$-quark spin is due to the orbital angular momenta of
  $u$, $d$ and $s$ quarks inside the constituent quark while the spin
 of current valence quark is screened by the spins of the quark--antiquark
 pairs.
It is also important to note the exact compensation between the spins
quark--antiquark pairs  and their angular orbital momenta:
\begin{equation}
\langle L_{\{\bar q q\}}\rangle= -J_{\{\bar q q\}}.\label{corr}
\end{equation} Since we consider effective lagrangian approach where gluon
 degrees of freedom are overintegrated, we do not discuss problems of the
principal separation and mixing of the quark orbital angular momentum and
 gluon effects in QCD (cf.  \cite{kiss}).  In the NJL--model
\cite{jaffe} the six-quark
 fermion operator simulates the effect of gluon operator
$\frac{\alpha_s}{2\pi}G^a_{\mu\nu}\tilde G^{\mu\nu}_a$, where $G_{\mu\nu}$ is
the gluon field tensor in QCD. The only effective degrees of freedom here are
quasiparticles; mesons and baryons are the bound states arising due to
residual interactions between the quasiparticles.

 Account for
axial anomaly in the framework of chiral quark models results in
 compensation of the valence quark helicity by  helicities of quarks
 from the cloud in the structure of constituent quark. The specific
 nonperturbative mechanism of such compensation is different in
 different approaches \cite{jaffe,fri},  e.g.
   the modification of the axial U(1) charge of
  constituent quark is considered to be generated by the interaction of
current quarks with flavor singlet field $\varphi^0$.
The apparent physical mechanism of such compensation has been discussed
recently in \cite{khar}.

  On these grounds we can
conclude that significant part of the spin of constituent quark should be
associated with the orbital angular momentum of quarks inside this
 constituent quark, i.e. the cloud quarks should rotate coherently inside
 constituent quark.

  The important point  what the origin of this orbital angular
  momentum is. It was proposed \cite{asy} to use an analogy with
 an  anisotropic extension of the theory of superconductivity
 which seems to match well with the above picture for a constituent
  quark.  The studies \cite{anders} of that theory show that the
   presence of anisotropy leads to axial symmetry of pairing
  correlations around the anisotropy direction $\hat{\vec{l}}$ and to
 the particle currents induced by the pairing correlations.  In
 another words it means that a particle of the condensed fluid
   is surrounded
 by a cloud of correlated particles ("hump") which rotate around
it with the
axis of rotation $\hat {\vec l}$.
(cf. Eq. (\ref{bal})
 Calculation of the
orbital momentum  shows that it is proportional to the density of the
correlated particles.
Thus, it is clear that there is a direct analogy
between  this picture and that describing the constituent quark. An
axis of anisotropy $\hat {\vec l}$ can be associated with the
polarization vector of  valence quark located at the origin of the
constituent quark.  The orbital angular momentum $\vec L$ lies
along $\hat {\vec l}$
(cf. Eq. (\ref{bal}).

 We argued that the existence of this orbital angular momentum, i.e.
 orbital motion of quark matter inside constituent quark, is the
 origin of the observed asymmetries in inclusive production at
  moderate and high transverse momenta.  Indeed, since the
 constituent quark has a small size
\[
r_Q=\xi/m_Q,\quad \xi\simeq 1/3,
 \quad m_Q\propto -\langle 0 | \bar q q| 0\rangle/\Lambda^2_\chi
\]
 the
asymmetry associated with internal structure of this quark will be
significant at $p_{\perp}>\Lambda_\chi\simeq 1$ GeV/c where interactions at
 short distances  give noticeable contribution.

 The behaviour of asymmetries in inclusive
 meson production was predicted \cite{asy} to have a corresponding
$p_{\perp}$--dependence, in particular, vanishing asymmetry at
$p_{\perp}<\Lambda_\chi $, its increase in the region of
$p_{\perp}\simeq\Lambda_\chi $, and $p_{\perp}$--independent
asymmetry at $p_{\perp}>\Lambda_\chi $.  Parameter
$\Lambda_\chi\simeq 1$ GeV/c is determined by the scale of chiral
symmetry spontaneous breaking.  Such a behaviour of asymmetry  follows
from the fact that the constituent quarks themselves have slow (if at
all) orbital motion and are in the $S$--state, but interactions with
$p_{\perp}>\Lambda_\chi $ resolve the internal structure of
constituent quark and ``feel'' the presence of internal orbital momenta
inside this constituent quark.

 It should be noted that at high $p_\perp$ we will  see
 the constituent quark being a cluster of partons
 which however should  preserve  their orbital
 momenta, i.e. the
 orbital angular momentum will be retained
 and the partons in the cluster are to be correlated.
 It should be stressed again that  a nonzero internal orbital
momentum of partons in the constituent quark means that there are
significant multiparton correlations.
Presence of such parton correlations
is in agreement with a high locality of strange sea in the nucleon.
 The concept of locality was proposed in \cite{ji} on the basis of analysis
of the recent CCFR data \cite{ccfr} for neutrino deep--inelastic scattering.
The locality serves as a measure of the local proximity of strange quark and
antiquark in momentum and coordinate spaces. It was shown \cite{ji} that the
CCFR data indicate that the strange quark and antiquark have very similar
distributions in momentum and coordinate spaces.

\section{Model for $\Lambda$--hyperon polarization}

 We  consider the hadron process of
 the  type \[ h_1 +h_2\rightarrow h_3^{\uparrow} +X\] with unpolarized
 beam and target. Usually we consider $h_1$ and $h_2$ being protons
 and $h_3$ --- $\Lambda$--hyperon.  Its polarization is being measured
through angular distribution of products in parity nonconserving $\Lambda$
decay.

   The picture of hadron consisting of constituent quarks embedded
 into quark condensate implies that overlapping and interaction of
 peripheral clouds   occur at the first stage of hadron interaction.
Under this, condensate  is
 being  excited  and  as  a result the quasiparticles , i.  e.
massive  quarks appear in the overlapping region.
It should be noted that
the condensate excitations are massive quarks,
since the vacuum is nonperturbative one and there is no overlap between
 the physical (nonperturbative) and bare (perturbative) vacuum
 \cite{isl,njl}.
 The  part of
 hadron  energy  carried  by  the  outer clouds of condensates being
released  in  the overlapping region, goes to the generation of
massive quarks.  Number of such quarks fluctuates.  The average
number of these  quarks in the framework of the geometrical
 picture can be estimated as follows:  \begin{equation} N(s,b)
\propto N(s)\cdot D^{h_1}_c\otimes D^{h_2}_c. \label{4} \end{equation}  Sign
$\otimes$ denotes convolution integral
\[
\int D_c^{h_1}(\vec{b'})D_c^{h_2}(\vec{b}-\vec{b'})d^2\vec{b'} .
\]
  The function $D^{h_i}_c$
describes condensate distribution inside hadron $h_i$ and $b$ is the
impact parameter of colliding hadrons $h_1$ and $h_2$.
  To estimate the
function $N(s)$ we can use the maximal possible value
$N(s)\propto\sqrt{s}$ \cite{trtu}.
Thus, as a result massive virtual quarks appear in the
overlapping region and  some mean field is generated.

Constituent quarks  located
 in the central part of hadron are supposed to scatter in a
 quasi-independent way by this mean field.

We propose the following mechanism for polarization of
$\Lambda$--hyperons based on the above picture for hadron structure.
Inclusive production
  of the hyperon $h_3$  results from two mechanisms: recombination of the
constituent quarks with virtual massive strange quark (low $p_{\perp}$'s,
soft interactions) into $h_3$ hyperon or from the scattering of a constituent
quark in the mean field, excitation of this constituent quark, appearence of
a strange quark as a result of decay of the constituent quark and subsequent
 fragmentation of strange quark in the hyperon $h_3$. The second mechanism is
determined by the interactions at distances smaller than constituent quark
radius and is associated therefore with hard interactions (high
$p_{\perp}$'s). This second mechanism could result from the single scattering
in the mean field, excitation and decay of constituent quark or from the
multiple scattering in this field with subsequent corresponding excitation
and decay of the constituent quark.  It is due to the multiple scattering by
mean field the parent constituent quark becomes polarized since it has a
nonzero mass \cite{szw} and this polarization results in polarization of
produced strange quarks and appearance of the corresponding  angular
orbital momentum.  Other mentioned mechanisms lead to production of
unpolarized $\Lambda$--hyperons.  Thus, we adopt a two--component picture of
hadron production which incorporates interactions at long and short
distances and it is the short distance dynamics which determines the
production of polarized $\Lambda$--hyperon.

It is necessary to note here, that after decay of the parent constituent
quark, current quarks appear in the nonperturbative vacuum and become a
quasiparticles due to the nonperturbative dressing with a cloud of
 $\bar q q$-pairs.  Mechanism of this process could be associated with the
strong coupling existing in the pseudoscalar channel \cite{khar,jaffe}.

Now we write down the explicit formulas for corresponding
inclusive cross--sections and polarization of hyperon $h_3$.  The following
expressions were obtained in \cite{tmf} which take into account
unitarity in the direct channel of reaction. They have the form
\begin{equation} \frac{d\sigma^{\uparrow,\downarrow}}{d\xi}=
8\pi\int_0^\infty bdb\frac{I^{\uparrow,\downarrow}(s,b,\xi)}
{|1-iU(s,b)|^2},\label{un} \end{equation} where $b$ is the impact
parameter of colliding hadrons. Here  function $U(s,b)$ is the
generalized reaction matrix (helicity nonflip one) which is
determined by dynamics of the elastic reaction
  \[ h_1+h_2\rightarrow h_1+h_2. \]
Arrows here denote the corresponding transverse polarization
of hyperon $h_3$.

The functions $I^{\uparrow,\downarrow}(s,b,\xi)$ are related to the
functions $U_n (s,b,\xi,\{\xi _{n-1}\})$ which are the multiparticle
analogs of the $U(s,b)$ and are determined  by  dynamics
of the exclusive processes \[ h_1+h_2\rightarrow
h_3^{\uparrow,\downarrow}+X_{n-1}.  \]
 The kinematical variables $\xi$ ($x$
and $p_\perp$, for example) describe the kinematical variables of the
 produced hyperon $h_3$ and the set of variables $\{\xi_{n-1}\}$ describe the
 system $X_{n-1}$ of $n-1$ particles.  It is useful to introduce the two
functions $I_+$ and $I_-$:  \begin{equation}
I_{\pm}(s,b,\xi)=I^\uparrow(s,b,\xi)\pm I^\downarrow(s,b,\xi),
\end{equation} where  $I_+(s,b,\xi)$ corresponds to unpolarized case.
The following sum rule takes place for the function $I_+(s,b,\xi)$:
\begin{equation} \int I_+(s,b,\xi)d\xi=\bar n(s,b)\mbox{Im}
U(s,b),\label{sr} \end{equation} where $\bar n(s,b)$ is the mean
multiplicity of secondary particles in the impact parameter
representation.

Polarization $P$  defined as the  ratio \[ P(s,\xi)=
\{\frac{d\sigma^\uparrow}{d\xi}-\frac{d\sigma^\downarrow}{d\xi}\}/
\{\frac{d\sigma^\uparrow}{d\xi}+\frac{d\sigma^\downarrow}{d\xi}\} \]
can be expressed in terms of the functions $I_{\pm}$ and $U$:
\begin{equation} P(s,\xi)={\int_0^\infty bdb
I_-(s,b,\xi)/|1-iU(s,b)|^2}/ {\int_0^\infty bdb
I_+(s,b,\xi)/|1-iU(s,b)|^2}.\label{xnn} \end{equation}

Using relations between transversely polarized states
$|\uparrow,\downarrow\rangle$ and helicity states $|\pm\rangle$,
  one can write down expressions for  $I_+$
and $I_-$ through
 the helicity functions
$U_{\{\lambda_i\}}$:
\begin{equation}
I_+(s,b,\xi)  =   \sum_{n,\lambda_1,
\lambda_2,\lambda_3,\lambda_{X_{n-1}}} n \int d\Gamma_n'
 |U_{n,\lambda_1,\lambda_2,\lambda_3, \lambda_{X_{n-1}}} (s,b,\xi,
\{\xi_{n-1}\})|^2,
\end{equation}
\begin{eqnarray}
 I_-(s,b,\xi)  =  & & \sum_{n,
\lambda_1,\lambda_2,\lambda_{X_{n-1}}} 2n  \int  d\Gamma_{n-1} \mbox{Im}[
 U_{n,\lambda_1,\lambda_2,+, \lambda_{X_{n-1}}}
 (s,b,\xi, \{\xi_{n-1}\})\nonumber\\
 & & U^*_{n,\lambda_1,\lambda_2, -, \lambda_{X_{n-1}}} (s,b,\xi,
\{\xi_{n-1}\})].\label{imi}  \end{eqnarray} Here the $\lambda_{X_{n-1}}$
denotes the set of helicities of particles from $X_{n-1}$ system; note that
in general this system as a whole has no definite spin or helicity.

Since in the model constituent quarks are quasi--independent ones
and the production of hyperon $h_3$ is the result of interaction
of one of them with the mean field, we can write the helicity functions
$U_{\{\lambda_i\}}$ as a sum
$U_{\{\lambda_i\}}=\sum_jU^{Q_j}_{\{\lambda_i\}}$
or simply as
$U_{\{\lambda_i\}}=NU^{Q}_{\{\lambda_i\}}$
taking into account that there are no constituent strange quarks among
the $N$ initial quarks in the colliding hadrons $h_1$ and $h_2$
(we do not consider here the processes with initial hadrons
 containing strange quarks and therefore all constituent quarks are
considered to be equivalent in respect to the production of the hyperon
$h_3$).  Superscript $Q$ denotes that the helicity function
$U^{Q}_{\{\lambda_i\}}$ describes the production of hyperon $h_3$ as
a result of interaction a quark $Q$ with the mean field.

In the model the spin--independent part
$I_+^Q(s,b,\xi)$
(note that $I_\pm(s,b,\xi)=N^2 I_\pm^{Q}(s,b,\xi)$)
gets contribution from the processes at small (hard
processes) as well as at large (soft processes) distances, i.e.
\[I_+^Q(s,b,\xi)= I^{hQ}_+(s,b,\xi)+ I^{sQ}_+(s,b,\xi),\] while the
spin--dependent part $I_-^Q(s,b,\xi)$ gets contribution from the interactions
at short distances only \[I_-^Q(s,b,\xi)=I^{hQ}_-(s,b,\xi).\]

  The presence of internal orbital momenta  in the structure
of constituent quark will lead to
  a certain shift in transverse momenta of produced hyperon,
 i.e. $p_\perp\rightarrow p_\perp\pm k_\perp$.
We suppose on the basis of Eq. (\ref{corr}) that there is a particular
flavor
compensation between spin and orbital momentum of strange quarks
inside constituent quarks, i.e.
\begin{equation}
L_{s/Q}=
-J_{s/Q}.\label{cor1}
\end{equation}
It seems to be a natural assumption and due to this
 the effect of shift of transverse momenta and polarization of
$\Lambda$--hyperon are directly connected
since the spin and polarization of $\Lambda$--hyperon
are  completely determined by those of the strange quark in the simple
 $SU(6)$ scheme. Eq. (\ref{cor1}) is quite similar to the conclusion made in
the framework of the Lund model \cite{and} but has different dynamical origin
 rooted in the mechanism of the spontaneous breaking of chiral symmetry.

 In the region of rather high
 transverse momenta $p_\perp>\Lambda_\chi$,
 the effect of this shift will be reduced to the phase factor in impact
 parameter representation \cite{asy}.
  Taking into account that quark matter distribution inside constituent
 quark has radius $r_Q$
and making the numerical estimation
$ k_{\perp s/ Q}=L_{ s/ Q}/r_Q$ we use  the following relation
on the grounds of considerations given in \cite{asy}:
\begin{equation} I_-^{hQ}(s,b,\xi)= \sin[\pm L_{s/Q}]
 I^{hQ}_+(s,b,\xi). \label{rel}
\end{equation}
Note that the sign is determined by the direction of rotation
of quark-antiquark pairs inside the constituent quark and
 since the value of  orbital angular momentum of $\bar s s$ quarks
in the constituent quark $ Q$ is proportional to the magnitude of its
polarization and mean orbital momentum of quarks in the constituent
quark, we can rewrite this relation in the form
\begin{equation} I_-^{hQ}(s,b,\xi)= \sin[
 {\cal{P}}_{Q}(x)
\alpha\langle L_{\{\bar q q\}}\rangle]
 I^{hQ}_+(s,b,\xi), \end{equation}
where
${\cal{P}}_{Q}(x)$ is the polarization of the
constituent quark $Q$ which is arising due to multiple scattering
in the mean field and
$\langle L_{\{\bar q q\}}\rangle$ is the mean value of internal angular
momentum inside the constituent quark.
Note that we consider the behaviour of polarization in the
 fragmentation region (where $x_F\simeq x$) and have taken the value of
 $L_{s/Q}$ to be proportional to
$\langle L_{\{\bar q q\}}\rangle$.

Thus, in this model polarization of strange quark is a result of multiple
scattering of parent constituent quark, correlation between the
polarization of strange quark and polarization of the constituent quark
and local compensation of spin and orbital angular momentum of
strange quark (cf. Eq. (\ref{cor1})).
The nonzero orbital angular momentum
leads to the shift in the transverse momentum of $s$--quark and
  produced
$\Lambda$-hyperon. This is the reason for the appearance of the factor
 $\sin[\pm L_{s/Q}]$ in Eq. (\ref{rel}).

The $x$--dependencies of the functions $I_+^{sQ}(s,b,\xi)$ and
$I_+^{hQ}(s,b,\xi)$ are determined by the distribution of constituent
quarks in  hadrons and by the structure function of constituent quark
respectively \cite{asy}:  \begin{equation} I_+^{sQ}(s,b,\xi)  \propto
\frac{1}{2}(\omega_{Q/h_1}(x)+\omega_{Q/h_2}(x))\Phi^{sQ}(s,b,p_{\perp})
\end{equation}
and
\begin{equation}
I_+^{hQ}(s,b,\xi)\propto \omega_{ s/Q}(x)
\Phi^{hQ}(s,b,p_{\perp}).  \end{equation}
Taking into account the above
relations, we can represent the polarization  $P$ in the form:
\begin{equation} P(s,x,p_{\perp})= \sin[{\cal{P}}_{Q}(x)
\alpha\langle L_{\{\bar q q\}}\rangle] {W_+^{hQ}(s,\xi)}/ {[W_+^{sQ} (s,\xi)
+W_+^{hQ}(s,\xi)]},\label{an} \end{equation} where the  functions $
W_+^{s,hQ}$ are determined by the interactions at long (s) and short (h)
distances:  \[ W_+^{s,hQ}(s,\xi)=\int_0^\infty
bdb{I_+^{s,hQ}(s,b,\xi)}/ {|1-iU(s,b)|^2}.  \]

\section{Behaviour of $\Lambda$--polarization}

As it has been already noted we consider the most simple case
of $\Lambda$--hyperon production. In this case  spin and
polarization of hyperon $h_3$ is completely determined by
the spin and polarization of $s$-quark from the internal
structure of parent constituent quark. The latter acquires
its polarization due to multiple scattering in the mean field.
This polarization is negative, e.g. in gluon external field it is
 \cite{szw}
\begin{equation}
 {\cal{P}}_Q\propto -I\frac{m_Qg^2}{\sqrt{s}}. \label{xpl}
\end{equation}
 It could have
significant value
since constituent quark in our case has a nonzero mass $m_Q\sim m_h/3$ and
 intensity of the mean field in the  model
 $I\sim\sqrt{s}$ since it is generated by the quasiparticles
whose average number is rising with energy like $\sqrt{s}$ \cite{trtu}.
Note that $g$ in Eq. (\ref{xpl}) is the
coupling constant of quark interaction with  external field.

 Thus on the basis of above considerations we take an  assumption that the
polarization of constituent quark is  energy independent and it is
approaching the maximal value $-1$ at $x=1$.  The assumption about maximality
of polarization at the constituent level has been made on the basis of recent
data of ALEPH collaboration \cite{alph} which made such indication in the
analysis of $\Lambda_b$ polarization in $e^+e^-$ interaction.

 We
 take also the simplest possible $x$--dependence of ${\cal{P}}_Q(x)$, i.e.
 the linear one:  \begin{equation} {\cal{P}}_{Q}(x)= {\cal{P}}_{Q}^{max}x
\end{equation} where ${\cal{P}}_Q^{max}=-1$.

 The behaviour of $\Lambda$--polarization in the model has a
significantly different $x$ and $p_\perp$-dependencies
in the regions of small and large
transverse momenta $p_{\perp}\leq \Lambda_\chi$ and $p_{\perp}\geq
\Lambda_\chi$. It is convenient to introduce the ratio
\[
R(s,\xi)=\frac{W_+^h(s,\xi)}{W_+^s(s,\xi)}=
\frac{2\omega_{s/Q}(x)}
{\omega_{Q/h_1}(x)+
\omega_{Q/h_2}(x)}
r(s,p_{\perp}), \] where
the function $r(s,p_{\perp})$ in its turn is the $x$--independent
ratio \[ r(s,p_{\perp}) =\frac{\int_0^\infty bdb
\Phi^h(s,b,p_{\perp})/|1-iU(s,b)|^2} {\int_0^\infty bdb
\Phi^s(s,b,p_{\perp})/|1-iU(s,b)|^2}.  \] The expression for the
polarization can be rewritten in the form \begin{equation}
 P(s,x,p_{\perp})= \sin[{\cal{P}}_{Q}(x)\alpha
\langle L_{\{\bar q q\}}\rangle] {R(s,x,p_{\perp})}/ {[1 +R(s,x,p_{\perp})]},
\label{pl}
\end{equation} The function $R(s,x,p_{\perp})\gg 1$ at
$p_{\perp}>\Lambda_\chi$ since in this region dominate short distance
processes  and due to the similar reason $R(s,x,p_{\perp})\ll 1$ at
$p_{\perp}\leq\Lambda_\chi$.  Thus we have simple
$p_{\perp}$--independent expression for polarization at
$p_{\perp}>\Lambda_\chi$ \begin{equation} P(s,x,p_{\perp})\simeq
 \sin[{\cal{P}}_{Q}(x)\alpha\langle L_{\{\bar q q\}}\rangle]
\label{las} \end{equation} and a more complicated one for
the region $p_{\perp}\leq\Lambda_\chi$ \begin{equation}
 P(s,x,p_{\perp})\simeq\sin[ {\cal{P}}_{\tilde Q}(x) \langle
L_{\{\bar q q\}}\rangle]
\frac{2\omega_{s/Q}(x)}
{\omega_{Q/h_1}(x)+
\omega_{Q/h_2}(x)} r(s,p_{\perp}). \label{sas}
\end{equation} As it is clearly seen from Eq. (\ref{sas}) the polarization at
$p_{\perp}\leq\Lambda_\chi$ has a nontrivial $p_{\perp}$--dependence.
In this region polarization vanishes at small $p_{\perp}$ and is
 also suppressed by the factor
${2\omega_{s/Q}(x)}/
{(\omega_{Q/h_1}(x)+
\omega_{Q/h_2}(x))}$,
which can be considered as the ratio
of sea and valence quark distributions in hadron.
The $x$--dependence of polarization in this kinematical region strongly
 depends on particular parameterization of these distributions.
However this dependence in
  the region of transverse momenta
 $p_{\perp}>\Lambda_\chi$  has
 a simple form reflecting corresponding dependence
 constituent quark polarization.
The curve for polarization at $p_\perp>\Lambda_\chi$ corresponding
 to the linear dependence of
${\cal{P}}_{Q}(x)$
 is presented in Fig. 1.
  The  value of
 $\langle L_{\{\bar q q\}}\rangle\simeq 0.4$ has been taken \cite{asy} on the
 basis of the analysis \cite{vos} of the DIS experimental  data.
To get agreement with experimental data we take the value of parameter
 $\alpha=0.8$.
 Using the above value
 of quark angular orbital momentum we obtain a good agreement with
 the data in the case of linear dependence of constituent quark polarization.
 Note that here we have assumed that spin structure of transversely polarized
constituent quark is the same as the spin structure of longitudinally
polarized constituent quark.

Qualitative $p_\perp$  dependence of polarization described above also is in
good agreement with corresponding experimental data.
To describe quantitevely the $p_\perp$ dependence
 of $\Lambda$--polarization, in particular, in the region
$p_\perp\leq\Lambda_\chi$ we should chose
an explicit parameterization of the cross--section ratio
 $R(s,x,p_\perp)$ for the hard and soft processes.  For that purpose
we can consider the simplest
parameterization of the function $R$ \begin{equation} R(s,x,
p_\perp)=C(x){\exp(p_\perp/m)}/{(p_\perp^2+\Lambda_\chi^2)^2}.  \label{rat}
\end{equation}
Such parameterization implies typical behaviour of cross--sections of
soft (exponential) and hard (power-like) processes. We take $m=0.2$
GeV which sets the scale of soft interactions at 1 fm and
$\Lambda_\chi=1$ GeV/c.
As an example we  consider data at $x=0.44$ which cover  wide range
of $p_\perp$'s.
The magnitude of $C(x)$ at the above value of x is
chosen to be 0.2 to get an agreement with the experimental data. The
corresponding curve and experimental data are given in Fig. 2 and as it can
be easily seen agreement with experiment is good.

\section{Conclusion and discussion}

Now we  summarize the main results of the considered model:
\begin{itemize} \item polarization of
$\Lambda$ -- hyperons arises as a result of the  internal structure of the
constituent quark and its multiple scattering in the mean field. It is
proportional to the orbital angular momentum of strange quarks initially
confined in the
constituent quark;
\item sign of polarization and its value are proportional to
polarization of the constituent quark gained due to the multiple
scattering in the mean field.
\end{itemize}

The main role in the model  belongs to the orbital angular
momentum of $\bar q q$--pairs inside the constituent quark while
constituent quarks themselves have very slow (if at all) orbital
motion and may be described approximately by $S$-state of the
 hadron wave function.  The observed $p_{\perp}$--dependence of
$\Lambda$--hyperon polarization  in inclusive processes seems
to confirm such conclusions, since
it  appears to show up beyond
$p_{\perp}>1$
GeV/c, i.e.  the scale where internal structure of  constituent
quark can be probed. Note, that short--distance interaction in this
approach observes coherent rotation of correlated  $\bar q q$--pairs
inside the constituent quark and not a gas of free  partons.

We have considered  the most simple case of $\Lambda$--hyperon
polarization. As a whole problem, the case of hyperon polarization is
extremely complicated and many reactions we did not attempt to account and
many questions are left unanswered.  However, few comments on the other
reactions and the underlying mechanism we could make. First, we would like to
note that experimental data show that proton polarization in inclusive
process $pp\rightarrow pX$ is zero. This fact can easily be understood in the
model.  Indeed, multiple scattering of constituent quarks in the mean field
has a lower probability compare to single scattering. Single scattering does
not polarize quarks and protons appear unpolarized in the
final state since
single scattering is dominant in this process. On the other hand multiple
scattering, excitation and decay of constituent quarks are correlated
mechanisms, that is the reason of $\Lambda$--hyperon polarization in the
model. Of course, $\bar s$-quarks also will be produced polarized, but
contrary to $s$-quark, which can easily recombine with constituent quarks of
parent protons to produce $\Lambda$, $\bar s$-quark has no such  possibility
and should pick up virtual massive quarks generated at the condensate
interaction.  Since polarization of produced $\bar\Lambda$--hyperons in the
process $pp\rightarrow\bar\Lambda X$ is almost zero we should conclude that
latter mechanism implies strong depolarization dynamics.  Thus we have to
suppose different mechanisms of $\Lambda$ and $\bar\Lambda$ formation at
final state. Those mechanisms have comparable strength at $x=0$, but
$\bar\Lambda$-production has to be suppressed at large $x$ in agreement with
the experimental data \cite{hell}.  To describe very different behaviour of
polarization in other hyperon production it seems that we need very detailed
knowledge of fragmentation dynamics \cite{xxx} which is unattainable at the
moment.

\section*{Acknowledgements}
We would like to thank J. Ellis, P. Galumian, D. Kharzeev and V. Petrov
 for useful discussions.

\newpage
\section*{Figure captions}
\bf Fig. 1 \rm The $x$--dependence of $\Lambda$--hyperon polarization
in the process $pp\rightarrow\Lambda X$ at $p_L=400$ GeV/c.\\[2ex]
\bf Fig. 2 \rm The $p_\perp$--dependence of $\Lambda$--hyperon polarization
in the process $pp\rightarrow\Lambda X$ at $p_L=400$ GeV/c.
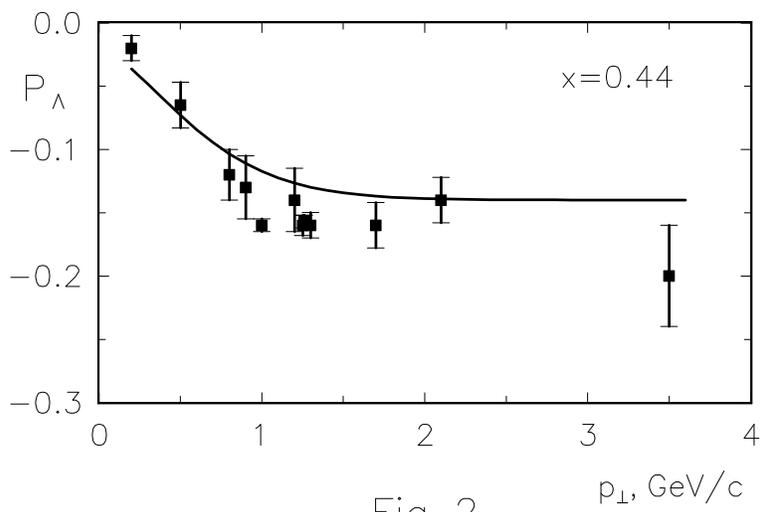
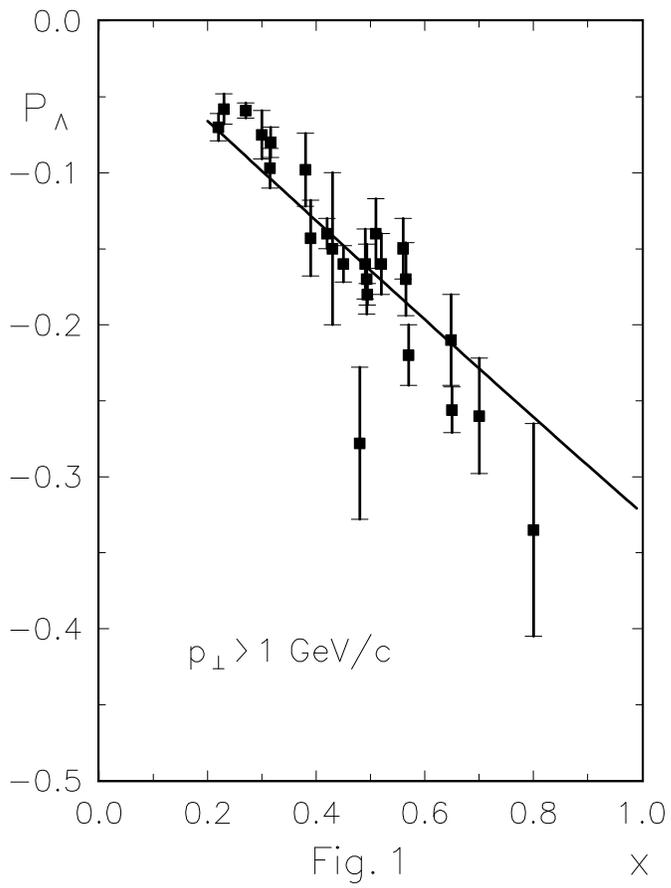
\end{document}